# Bright-Field AAPSM Conflict Detection and Correction

Chiang, C. Synopsys
Kahng, A. University of California, San Diego
Sinha, S. Synopsys
Xu, X.* University of California, San Diego
Zelikovsky, A. Georgia State University

## Abstract

As feature sizes shrink, it will be necessary to use AAPSM (Alternating-Aperture Phase Shift Masking) to image critical features, especially on the polysilicon layer. This imposes additional constraints on the layouts beyond traditional design rules. Of particular note is the requirement that all critical features be flanked by opposite-phase shifters, while the *shifters* obey minimum width and spacing requirements. A layout is called phase-assignable if it satisfies this requirement. If a layout is not phase-assignable, the phase conflicts have to be removed to enable the use of AAPSM for the layout. Previous *work* has sought to detect a suitable set of phase Conflicts to be removed, as well as correct them *[3, 4,* 5, 6, *8]*.

The contributions of this paper are the following: (1) a new approach to detect a minimal set of phase conflicts (also referred to as AAPSM conflicts), which when corrected will produce a phase-assignable layout; (2) a *novel* layout modification scheme for correcting these AAPSM conflicts. The proposed approach for conflict detection shows significant improvements in the quality of results and runtime for real industrial circuits, when compared to previous methods. To the best of our knowledge, this is the first time layout modification results are presented for bright-field AAPSM. *Our* experiments show that the percentage area increase for making a layout phase-assignable ranges from *0.7-11.8%*.

## 1 Introduction

Alternating-Aperture Phase Shift Masking (AAPSM) is a form of strong RET which will be widely used for imaging the polysilicon layer at the sub-100nm nodes. There are several variants of AAPSM. In this paper, we will mainly focus on **Bright-Field** AAPSM. This is the variant that is most likely to be used for the polysilicon layer. In Bright-Field AAPSM, a critical feature is a shape in the design such as a wire. Each critical feature is flanked by two phase shapes or shifters, which must be assigned opposing phases in order to create destructive interference between them. These shifters must also satisfy given design-rule (i.e., spacing and width) constraints.

This requirement gives rise to the **phase assignment problem** in layouts, which can be stated as follows: ***Given a layout with shifters inserted around each critical feature*[1]*, find a phase-assignment solution such that the following conditions are satisfied:***

1. Condition *1:* Shifters on opposite sides of *every* critical feature are assigned opposite phases *(0° and 180°)*; and
2. Condition 2: Shifters that are separated by less than the minimum shifter spacing should be merged and assigned the same phase. Henceforth, two shifters separated by less than the minimum shifter spacing will be referred to as **overlapping shifters.**

Figure 1 illustrates an example where the above conditions are violated due to a non-localized cyclic sequence of phase-shifters that cannot be properly mapped. We say two adjacent phase-shifters are in ***AAPSM conflict*** if they belong to such a cycle of phase dependencies. Previous work on eliminating these cycles falls into two major categories. Modifying the layout by increasing spacing between features or widening critical features is one approach [1, 2, 4, 3, 5, 6]. Most of these works focus on dark-field AAPSM. These methods will be discussed in more detail in the next section. The other method for eliminating AAPSM conflicts is to modify the mask instead of the layout. In this approach, phase conflicts are corrected by introducing cuts in the shifters at points where two opposite-phase shifters meet [8]. While this approach avoids the area increase that can arise from layout modification, it can significantly complicate the mask creation process, which may limit its use.

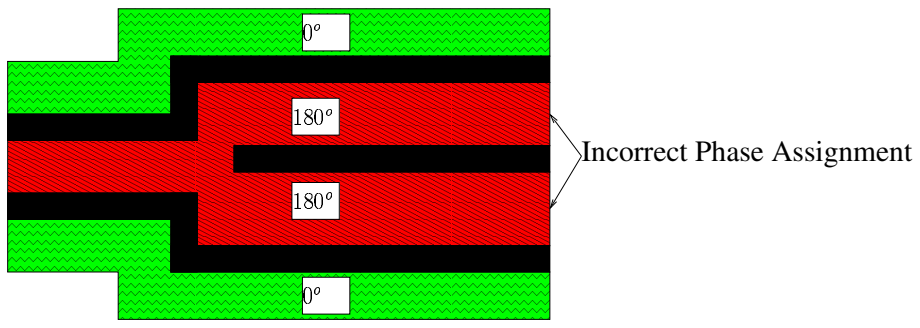

Figure 1: **Example of incorrect phase assignment.**

The key contributions of this paper are the following: (1) a new scheme for detecting a minimal number of AAPSM conflicts, that when corrected will produce a ***phase-assignable*** layout; and *(2)* a novel layout modification scheme for correcting these AAPSM conflicts. The proposed scheme for conflict detection shows significant improvements in quality of results and runtime compared to previous methods. To the best of our knowledge, this is the first time layout modification results are presented for bright-field AAPSM. In Section 2, we give a detailed description of the previous work on detecting a minimum set of AAPSM conflicts that when corrected in the given layout will make it ***phase-assignable*** and the work on layout modification for dark-field AAPSM. The conflict detection work for bright-field AAPSM was done based on

---

***This work was done during the course of his internship at ATG, Synopsys.**

**[1]Typically, a feature whose width is below a certain threshold is a critical feature.**

the assumptions that shifter widths are not larger than half the minimum feature length and that only minimum-width features need to be phase-shifted. These assumptions may no longer be valid for smaller technology nodes. In this paper, we relax these assumptions and present a new flow to detect and correct AAPSM conflicts. Section 3 describes the general flow of our approach. In Section 3.1, we describe our scheme for selecting a minimal number of AAPSM conflicts for correction, and outline its advantages. Section 3.2 discusses a novel layout modification scheme for correcting the AAPSM conflicts chosen by the detection algorithm. We present experimental results in Section 4 and end with conclusions and directions for future work in Section 5.

# 2 Related Work

The phase assignment problem was first posed by Moniwa et al. [1] and Ooi et al. [2]. The authors suggested methods for detecting cases when there are no valid phase assignments. Moniwa et al. [4]proposed an approach to enumerate all odd cycles and iteratively delete edges by increasing the lower bound of separation between corresponding features. Ooi et al. [3] suggested a compaction-based method to transform a mask layout into a ***phase-assignable*** one using an initial phase assignment of the mask layout. Both approaches were given in the context of dark-field AAPSM, and suffer from the common problem that they did not attempt to minimize the modification of the original layout.

The work in [5, 6] looks at optimal algorithms for phase assignment while minimizing the amount of layout modification. The phase assignment problem on the layout is converted to a graph bipartization problem on a graph constructed from the layout (called the ***conflict graph*** for dark- field AAPSM [5] and the ***feature graph*** for bright-field AAPSM [6]). It is proved that the layout is ***phase-assignable*** if and only if the corresponding graph is bipartite. The graph edges are weighted to provide a measure of the layout impact caused by increasing the spacing between corresponding features. The authors [5] formulate a minimum-weight bipartization problem (referred to as the minimum distortion problem in [5]) on the graph $G = (V,E)$ constructed from the layout; this formulation seeks the minimum weight edge set $M$ such that the graph $(V,E - M)$ is bipartite. An optimal solution is presented for the case where G is an embedded planar graph[2]. The optimal bipartization algorithm, henceforth referred to as ***Bipartize,*** works as follows: First, the geometric dual $D$ of G is constructed. This step requires that the input graph be an embedded planar graph. The authors prove that solving the minimum-weight bipartization problem on G is equivalent to solving the ***optimal T-join problem***[3] on the dual graph $D$, where $T$ is the set containing **all** the odd degree nodes in $D$. The ***optimal T-join problem*** of $D$ can be reduced to a minimum-weight perfect matching on a new graph $D'$. The graph $D'$ is constructed using a set of ***optimized gadgets*** proposed in [5].

The set M represents the AAPSM conflicts that when corrected will produce a ***phase-assignable*** layout. For dark-field AAPSM, the authors in [6] present results on area increases for re-compacting the layout using a "one-shot" compaction flow proposed in [7]. However, no results for layout modification are presented for bright-field AAPSM. Furthermore, the implicit assumption that the feature graph is an embedded planar graph does not necessarily hold true for sub-100 nm technology nodes: this indeed affects the optimality of the edge deletion solution, as our experiments below indicate. It is proved in [6] that the feature graph is an embedded planar graph if the maximum shifter dimension is less than half the minimum feature length. This assumption does not take into account the fact that shifters of opposite phase must be separated by a certain distance, and that features larger than the minimum width may also need to be phase-shifted. In the presence of these additional requirements, experimental evidence shows that the feature graph may no longer be an embedded planar graph.

[2] **An embedded planar graph is one that has no line crossings when when embedded in *a* plane.**

[3] **The optimal T-join problem of a graph seeks a minimum- weight edge set *A* such that a node *u* is incident to an odd number of edges of A if and only if *u* belongs to the node subset *T* of the given graph.**

# 3 Proposed Flow

The proposed approach for AAPSM conflict detection and correction is presented below:

1. ***Reduction of Phase Assignment Problem to Embedded Planar Graph Bipartization.***
   (a) **Phase Conflict Graph Generation. Given a layout $L$, a phase conflict graph $G$ is constructed from it. Like the feature graph in [6], it has the property that it is bipartite if and only if $L$ is phase-assignable. However, it affords advantages over the feature graph, as we will discuss in the next section.**
   (b) **Planar Graph Embedding. The phase conflict graph $G$ is not necessarily an embedded planar graph, though in practice it has a much smaller number of line crossings than the planar embedding of the feature graph. However, the optimal bipartization algorithm requires an embedded planar graph. Hence, G is converted to an embedded planar graph $G_P$ by applying the planar embedding algorithm in [5] and greedily removing minimum weight edges that cross other edges. These edges are added to a potential set of AAPSM conflicts $P$.**

2. ***Embedded Planar Graph Bipartization.*** A modified version of the minimum-weight bipartization algorithm ***Bipartize,*** described in Section 2, is applied to $G_P$. The main difference between **our** version of the optimal bipartization algorithm and ***Bipartize*** is that we use a different reduction of the T-join problem to minimum-weight perfect matching; this is described in further detail in Section 3.1.2. Let $D$ denote **a minimal set of AAPSM conflicts** which when removed will ensure that G is bipartite, and $L$ is ***phase-assignable.*** The list of edges deleted by the optimal bipartization algorithm is added to $D$. If the phase conflict graph G is an embedded planar graph, $D$ also denotes the minimum set of AAPSM conflicts that when corrected will produce a ***phase-assignable*** layout.

3. ***Computation of final set of AAPSM conflicts.*** It is necessary to check if any of the edges deleted during planar embedding, i.e., the edges in $P$, belong to odd cycles. This is accomplished by coloring $G_P$ with two colors such that only the endpoints of the edges in D have the same color. If the endpoints of an edge $e \in P$ have the same color, then $e$ violates the bipartition condition of G and hence is added to the set D. At this point, D has a minimal set of edges/AAPSM conflicts which when removed will make G bipartite.
4. ***Layout Modification.*** In this step, the layout is modified by adding end-to-end horizontal and/or vertical spaces throughout the layout such that all the AAPSM conflicts in D are corrected. An AAPSM conflict is corrected if the shifters that are represented by the endpoints of the edge are separated to satisfy the shifter spacing rules. It is not sufficient to only increase the spacing between the shifters corresponding to the endpoints of the edge. This might cause DRC violations elsewhere and may need an additional re-compaction step. End-to-end spaces added through the layout will avoid introducing DRC errors since the additional space is added uniformly through the layout.

## 3.1 AAPSM Conflict Detection Scheme

We now discuss the two key features of our AAPSM conflict detection scheme, the phase conflict graph and a new reduction to solve the T-join problem.

### 3.1.1 Phase Conflict Graph

The layout is assumed to be composed of a set of non-overlapping rectangles. Given a layout $L$, the ***phase conflict*** graph $G = (ES \cup O, E)$ consists of two types of nodes $ES$ and $O$ and edges $E$:

1. For each shifter, create an edge shifter node $es \in ES$.
2. For two overlapping shifters $s_1$ and $s_2$, create an overlap node $o \in O$ and place it at the center of the line connecting $es_1$ to $es_2$ (here $es_1$ and $es_2$ represent the edge shifter nodes for $s_1$ and $s_2$, respectively). Create two edges $e_1 = (es_1, o)$ and $e_2 = (o, es_2)$ and include them in $E$.
3. Create an edge $e = (es_1, es_2)$ between the two shifters that are on opposite sides of a critical feature.

Note that an edge in the phase conflict graph always represents a pair of shifters.

**Theorem 1** *Let G be a phase conflict graph corresponding to the layout L. The phase assignment problem has a feasible solution for L if and only if G is bipartite.*

**Proof:** Omitted for the sake of brevity. □

The phase conflict graph of a layout is shown in Figure 2(a). For reference, the feature graph for the same layout is shown in Figure 2(b). The key advantage of the phase conflict graph over the feature graph [6] is that it reduces the number of line crossings by representing a pair of overlapping shifters by a straight line between the corresponding edge shifter nodes. This is in contrast to the feature graph where the edges makes a detour to the geometric center of the region of overlap and hence increase their chances of crossing another edge. In most examples, the phase conflict graph also has a smaller number of nodes and edges than the feature graph. These two factors help reduce the number of line crossings when the phase conflict graph is used. Hence, the loss of optimality of our AAPSM conflict detection algorithm is greatly reduced. Experimental results on a large set of industrial benchmarks (presented in Section 4) show that the number of AAPSM conflicts that are chosen for correction is indeed substantially smaller when the phase conflict graph is used.

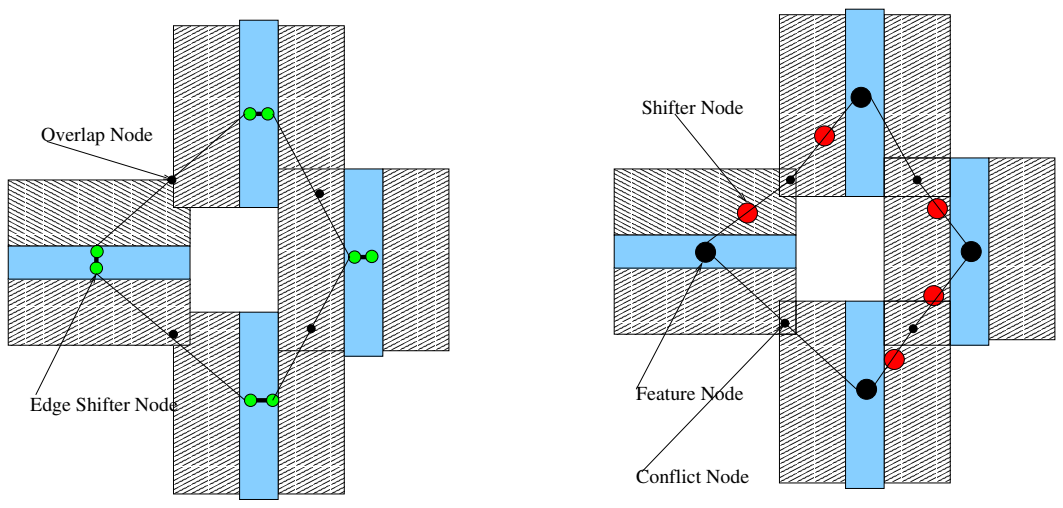

(a) Phase Conflict Graph (b) Feature Graph

Figure 2: Phase Conflict graph versus Feature graph for the same layout. The solid rectangles denote the features and the patterned rectangles denote the shifters.

### 3.1.2 Reduction via Generalized Gadget

As described earlier, the optimal bipartization problem can be reduced to the T-join problem, which in turn *can* be reduced to the minimum-weight perfect matching problem [5]. In this section, we present a new reduction from the T-join problem to the minimum-weight perfect matching problem. Our reduction uses a new set of gadgets called the ***generalized gadgets.***

Let the graph on which we want to solve the T-join problem be denoted as $G = (V, E)$. A ***generalized gadget graph,*** henceforth denoted as $G^G = (V', E')$, is constructed from G as follows:

1. Each edge $e = (v, v') \in G$ is assigned to one of its endpoints $v$ or $v'$. The assignment procedure is arbitrary with the only requirement being that each odd-degree node in G should be assigned an odd number of edges and each even-degree node should be assigned an even number of edges[4].
2. Each node $v \in V$ corresponds to a set of nodes in $G^G$. This set of nodes is the ***generalized gadget,*** denoted as $G_v$, of $v$. For each edge $e = (v, v')$ incident to $v$: if $e$ is assigned to $v$, a ***true*** node $t(v, e)$ is added to $G_v$, else a ***ghost*** node $g(v, e)$ is added to $G_v$.
3. The nodes in $G_v$ are connected to each other by weighted edges. The weights of the edges are computed as follows:
   (a) Weight is $0$ between two true nodes $t(v, e)$ and $t(v, e')$.
   (b) Weight is $w(e)$ between a ghost node $g(v, e)$ and a true node $t(v, e')$.
   (c) Weight is $w(e')$ between a true node $t(v, e)$ and a ghost node $g(v, e')$.
   (d) Weight is $(w(e) + w(e'))$ between two ghost nodes $g(v, e)$ and $g(v, e')$.
4. For each $e = (v, v') \in E$, the true node $t(v, e)$ and its corresponding ghost node $g(v', e)$ are connected to each other via 0-weight edges and a dummy node.

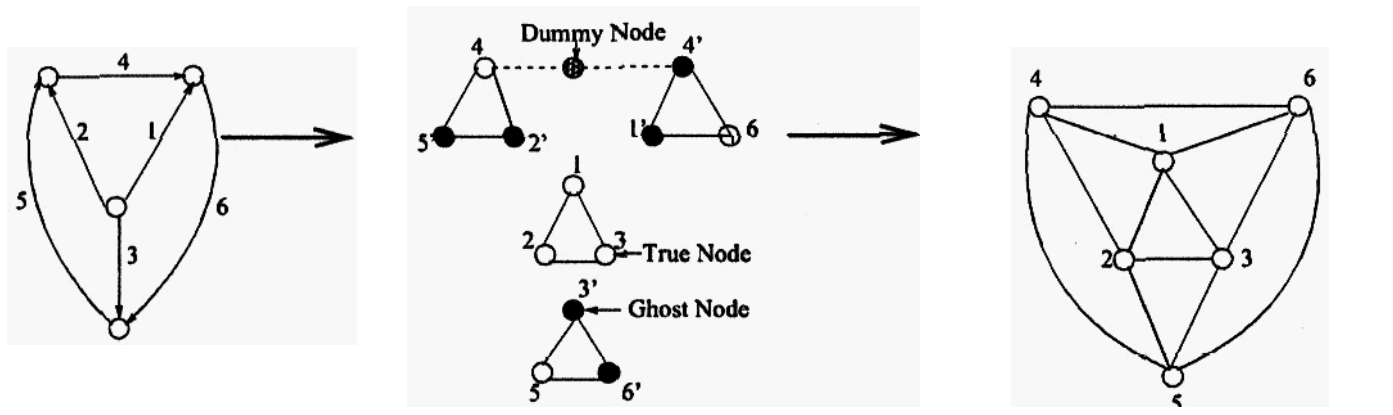

(a) Original Graph (b) Generalized Gadgets for each node (c) Generalized gadget graph

Figure 3: The original graph in (a) is converted to the generalized gadget graph (c). Dummy nodes between all true and ghost node pairs are not shown.

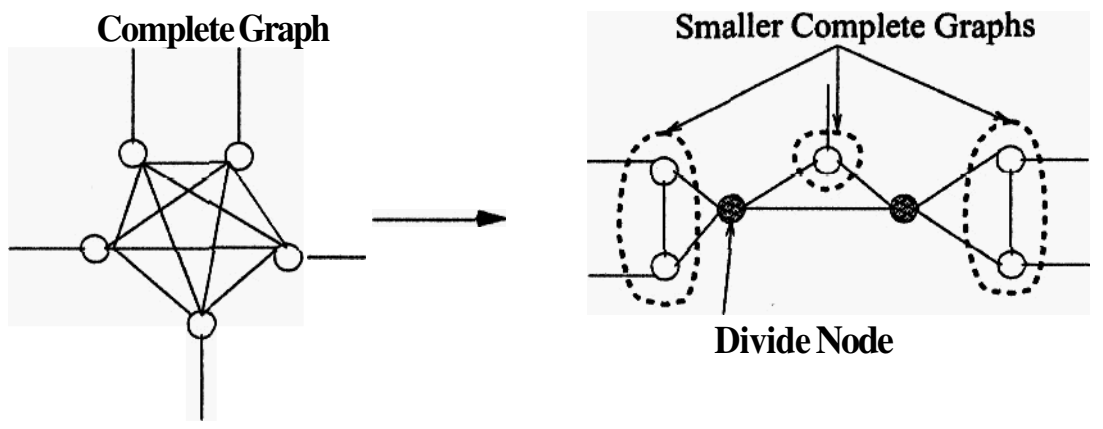

Figure 4: Modified Complete Gadget for node of degree 5.

It should be noted that in reality ghost nodes and the dummy nodes *are not explicitly represented* in the gadget graph. Thus a ghost node $g(v,e)$ in $G_{v'}$ is simply a pointer to the true node $t(v,e)$ in G,. The conversion of a graph G to a generalized gadget graph is illustrated in Figure 3.

**Theorem** 2 ***The Optimal T-join problem for a graph $G = (V,E,w,T)$, where T is the set containing all the odd-degree nodes in V, can be reduced to a minimum-weighted perfect matching problem on the generalized gadget graph $G' = (V',E',w')$.***

**Proof Sketch:** Omitted for the sake of brevity. □

In our implementation, the construction of the *generalized gadget* graph is further improved. Each *generalized gadget* is decomposed into $k$ smaller complete subgraphs connected together by $(k-1)$ divide nodes. The decomposition must satisfy the condition that the sum of nodes in the smaller complete subgraphs should be equal to the original number of nodes in the *generalized gadget.* The details of the construction are skipped but an example with a *generalized gadget* of size 5 is shown in Figure 4. It can also be proved that these improved gadgets provide a correct reduction. The *optimized gadgets* used for the reduction in [5] are a special case of these improved gadgets. The *optimized gadgets* only allow the complete sub-graphs to be of size less than equal to 3. In our gadgets, the complete sub-graphs can be of any size. This flexibility enables us to reduce the number of nodes in the gadget graph and produces a 16% improvement in runtime, when compared to the *optimized gadget.*

## 3.2 Layout Modification Scheme

The primary task of AAPSM-related layout modification is to correct the AAPSM conflicts that the conflict detection algorithm selected in the previous step. An AAPSM conflict is corrected by adding the required amount of space between the corresponding shifters such that they are separated by the required shifter spacing. But only increasing the spacing between the shifters corresponding to the AAPSM conflict may cause additional DRC violations. Hence, end-to-end spaces are added throughout the layout.

[4] It can be shown that this requirement can always be satisfied because an edge $e = (v, d)$ can be assigned to both $v$ and $d$ at the same time, if needed.

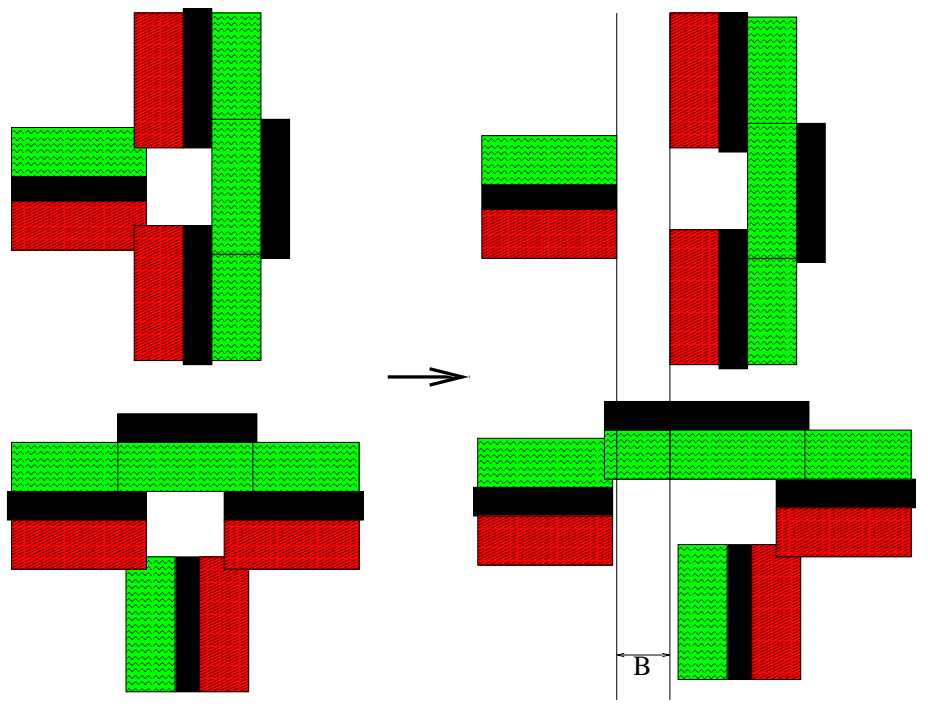

Figure 5: Inserting a vertical space of width B to remove multiple AAPSM conflicts

The basic idea of the proposed layout modification algorithm is illustrated in Figure 5. In this scheme, horizontal and vertical spaces of variable width are added throughout the layout to correct the chosen AAPSM conflicts. Note that adding these horizontal and vertical spaces cannot introduce any spacing violations since the spaces are added uniformly from one end of the layout to the other. Moreover, the spaces are added such that only the length of features are increased but the widths of the features remain the same. This prevents any major timing problems after layout modification. A similar layout modification scheme was briefly mentioned in [7] but no details were provided.

The problem for layout modification can be stated as follows: ***Given the set D of AAPSM conflicts that have to be corrected, determine the minimum number and the widths of the end-to-end horizontal and/or vertical spaces that need to be added.***

The proposed layout modification algorithm is given below:

1. For each AAPSM conflict, determine if it can be corrected by adding a vertical space or a horizontal space or both.
2. Create a set of intervals for each AAPSM conflict. If the conflict can be corrected by adding a horizontal (vertical) space, then its interval is the projection of the edge corresponding to the AAPSM conflict in the horizontal (vertical) axis. If it can be corrected by either, then it has two intervals: one horizontal and the other vertical.
3. Define a grid in the layout using the endpoints of the intervals. Each horizontal (vertical) grid-line specifies a position where a horizontal (vertical) space can be inserted.
4. Set up a weighted set covering problem. The elements of the universal set are the AAPSM conflicts in $D$. Each grid-line represents a subset, whose elements are the AAPSM conflicts that can be corrected by adding space at that grid-line. The weight of the set is equal to the largest amount of space needed to correct the conflicts that intersect the grid-line. The solution of the covering problem provides the locations and widths of the horizontal and vertical spaces that have to be added to the layout. A covering solver from Berkeley is used for our experiments [11].

| Design | # polygons/# Shifter Overlaps | # of AAPSM Conflicts (QoR) | | | | Matching Time in sec (CPU) | |
|---|---|---|---|---|---|---|---|
| | | NP | FG [6] | GB | PCG (proposed) | O-Gadget [5] | G-Gadget (proposed) |
| block1 | 1604/3720 | 62 | 318 | 230 | 141 | 0.20 | 0.22 |
| block2 | 3329/10669 | 258 | 394 | 513 | 259 | 5.73 | 5.15 |
| design1 | 10274/24580 | 665 | 1675 | 925 | 699 | 1.47 | 1.22 |
| design2 | 13630/32257 | 895 | 2204 | 1162 | 918 | 2.17 | 1.65 |
| design3 | 21868/53749 | 1574 | 3843 | 2185 | 1673 | 3.73 | 3.25 |
| design4 | 20425/50059 | 1396 | 3603 | 1877 | 1484 | 3.56 | 2.82 |
| design5 | 25784/63760 | 1857 | 4550 | 2586 | 1972 | 4.68 | 3.87 |
| design6 | 48787/157668 | 4925 | 15754 | 7497 | 6525 | 15.5 | 12.0 |
| design7 | 44121/142707 | 4592 | 12969 | 6445 | 5662 | 15.9 | 12.4 |
| design8 | 72101/237557 | 8067 | 24395 | 12034 | 10531 | 27.0 | 20.4 |
| design9 | 105882/376707 | 15290 | 39738 | 23693 | 18974 | 51.7 | 42.2 |
| FullChip | 159070/552767 | 23606 | 56881 | 36840 | 28843 | 79.6 | 66.6 |

Table 1: **AAPSM** conflict detection using the proposed enhancements

This approach is most suitable for **AAPSM** conflicts that cannot be corrected by adding simple DRC rules [9, 10]. This scheme could also be used to determine the best approach for correcting the selected **AAPSM** conflicts, i.e. to decide which conflicts are best corrected by layout modification and which conflicts are best corrected by mask splitting[6]. For instance, if a large number of **AAPSM** conflicts can be corrected by adding an end-to-end space at a single horizontal or vertical grid-line, it may make more sense to eliminate all of them using layout modification. Our experiments indicate that it is often possible to find a large number of such **AAPSM** conflicts. On the other hand, if the end-to-end space added to correct an **AAPSM** conflict does not correct too many other **AAPSM** conflicts, it may make sense to correct it using the mask splitting approach.

# 4 Experimental Results

In this section, we describe the experiments we conducted for testing the benefits of the proposed ideas. All our examples are ***90*** *nm* designs and assume typical values of threshold width for critical features, shifter dimensions and shifter spacing. We focus only on **AAPSM** conflicts that can be solved by increasing the space between features in the layout. Thus, **AAPSM** conflicts caused by T-shapes are not handled. These can be corrected by feature widening or mask splitting [8] (we are exploring extensions to our method to handle them as well). Even though, **AAPSM** conflicts caused by local line-end conflicts can be detected and corrected using our approach, they axe also not considered as they can be efficiently detected and corrected using additional DRC checks during layout generation [9, 10].

Table 1 compares the quality of results (number of edges deleted, or in other words, number of **AAPSM** conflicts selected for correction) and runtime for the proposed flow and other state-of-the-art approaches *[5,* 6]. We present results on two counts:

1. Quality of Results (# of **AAPSM** conflicts selected for correction): A smaller number of **AAPSM** conflicts is preferred for correction as this would minimize the amount of modification needed for the mask/layout.
   (a) Column *NP* specifies the number of **AAPSM** conflicts selected only by the optimal bipartization algorithm, when the **phase conflict *graph*** representation **of** the layout is used. (Step 1(a)-2 of the flow in Section 3).
   (b) Column FG specifies the number of **AAPSM** conflicts selected after including the cost of planar embedding to the optimal bipartization cost, when the **feature** graph representation **of** the layout is used (Steps 1(a)-3).
   (c) Column PCG specifies the number of **AAPSM** conflicts selected after including the cost of planar embedding to the optimal bipartization cost, when the **phase conflict graph** is used **for** representing the layout (Steps 1(a)-3)[6].
   (d) **Column** GB specifies the number of **AAPSM** conflicts selected using a **greedy bipartization** algorithm that does not require the input graph to be an embedded planar graph. A maximum weight spanning tree T is constructed by greedily selecting the largest weight edge at every step, until no more edges *can* be added. The leftover edges that cannot be included in T denote the **AAPSM** conflicts selected **for** correction.

The difference between the results of ***NP*** on one hand and FG and ***PCG*** on the other is due to the line crossings during the planar embedding of the feature graph and the phase conflict graph, respectively. Experiments on a wide range of industrial layouts show that the proposed flow (optimal bipartization of the phase conflict graph after planar embedding) consistently selects a smaller number of **AAPSM** conflicts for correction when compared to the previous approach (optimal bipartization of the feature graph after planar embedding), and in most cases is quite close to the solution that does not take the planar embedding cost into account (Column *NP*). This is due to the fact that the phase conflict graph produces a relatively small number of line crossings during planar embedding, which has the desirable effect of making the solution close to the optimal one. Comparison with a greedy bipartization algorithm (Column ***GB)*** shows that the proposed **AAPSM** conflict detection scheme selects a much smaller set of **AAPSM** conflicts for correction, in spite of the loss of optimality due to planar embedding. Thus, we can conclude that optimal bipaxtization is preferable for this application, in spite of the additional planar embedding step.

[5] It seems likely that a mix of both approaches will be used since both techniques have their advantages and disadvantages.

[6] It should be noted that both Steps (b) and (c) use the same planar embedding scheme.

| Design | Area | Conflict | Grid | Max | % Area Increase |
|---|---|---|---|---|---|
| cell1 | 48.83 | 5 | 5 | 1 | 4.9 |
| cell2 | 23.2 | 6 | 5 | 1 | 6.6 |
| cell3 | 78.55 | 10 | 4 | 3 | 1.9 |
| cell4 | 17.56 | 10 | 4 | 3 | 4.5 |
| cell5 | 24.71 | 8 | 5 | 3 | 3.0 |
| ip1 | 1017.47 | 11 | 9 | 2 | 3.9 |
| ip2 | 2119.8 | 19 | 11 | 4 | 2.0 |
| ip3 | 22631.53 | 150 | 38 | 9 | 0.7 |
| ip4 | 11255.01 | 272 | 90 | 64 | 4.1 |
| ip5 | 8887.48 | 323 | 129 | 3 | 1.5 |
| ip6 | 21887.6 | 561 | 356 | 4 | 3.8 |
| ip7 | 169700.34 | 1613 | 511 | 6 | 11.8 |

Table 2: Layout modification for a variety of designs.

2. Runtime:
   (a) Column *0-Gadget* denotes the runtime for matching when the *optimized gadgets* [5] are used for translating the T-join problem to the minimum-weight perfect matching problem.
   (b) Column *G-Gadget* denotes the runtime for matching when the *generalized gadgets* are used for the same.

   As far as the runtimes are concerned, we find that except for one small example, matching using the ***generalized* gadget** graph is much faster than matching using the **optimized gadget** graph (average runtime improvement is **16%**).

The proposed flow is also quite robust and could be used on a full-chip layout with approximately **160K** polygons (last example in Table 1).

Table 2 reports the results of using the proposed layout modification scheme for correcting the **AAPSM** conflicts chosen by the detection step on some example layouts. Column **Area** reports the area of the designs in square microns. Column ***Conflict*** specifies the number of **AAPSM** conflicts selected by the detection algorithm for each design. The number of grid-lines where end-to-end horizontal or vertical spaces are added is reported in Column *Grid*. The maximum number of **AAPSM** conflicts that can be removed by adding space at a grid-line is reported in Column *Max*. It should be noted that a considerable fraction of the **AAPSM** conflicts can be corrected by adding a single end-to-end space. The last column reports the percentage area increase for these layouts as a result of the added spaces. The area increase for these layouts ranges from 0.7-11.8%, with an average increase of **4%**. The experimental results demonstrate that this layout modification scheme is a viable option for correcting **AAPSM** conflicts. Furthermore, the small increase in area for handling **AAPSM** conflicts shows that layout modification itself can be a feasible option for dealing with **AAPSM** conflicts.

## 5 Conclusions

In this paper, we proposed a new scheme for detecting a minimal set of **AAPSM** conflicts in polysilicon layouts without making any assumptions on shifter widths and spacings. Experiments on a large set of industrial examples indicate that the proposed approach of representing the layout using a phase conflict graph and solving the T-join problem using **generalized gadgets** significantly reduces the number of **AAPSM** conflicts selected for correction and improves the runtime, respectively when compared to previous methods. One possible extension would be to investigate better planar embedding schemes as it seems highly unlikely that the graphs constructed from the layout will be embedded planar graphs for realistic shifter parameters. Further runtime improvements for the **AAPSM** conflict detection step are also currently being investigated.

A simple and efficient layout modification scheme was also proposed. The area increase for making a layout **phase-assignable** ranges between **0.7-11.8%**. This makes the proposed layout modification scheme an attractive scheme for correcting **AAPSM** conflicts after the layout is deemed DRC correct. We are currently investigating approaches to handle constraints imposed by multiple layers and also incorporate feature widening as an option for correcting **AAPSM** conflicts in **our** scheme. Finally, extensions of the layout modification scheme to handle standard-cell blocks, that can restrict the insertion of cuts to certain regions and exploit the white-space inherent in the layout, are being investigated.